\documentclass[12pt,onecolumn,english,technote,onecolumn]{IEEEtran}
\usepackage[T1]{fontenc}
\usepackage[utf8]{inputenc}
\usepackage{xcolor}
\usepackage{babel}
\usepackage{float}
\usepackage{amsmath}
\usepackage{amssymb}
\usepackage{graphicx}
\usepackage{setspace}
\usepackage{esint}
\usepackage[pdfusetitle,
 bookmarks=true,bookmarksnumbered=false,bookmarksopen=false,
 breaklinks=false,pdfborder={0 0 1},backref=false,colorlinks=true]
 {hyperref}
\hypersetup{
 pdfborderstyle=,linkcolor=blue,citecolor=blue,urlcolor=blue}

\makeatletter
\usepackage[font=normalsize]{caption}
\usepackage[noadjust]{cite}
\usepackage{varwidth}

\makeatother

\begin{document}
\title{An Integral Equation for the Surface Current in the Problem of the
Natural Modes of a Waveguide Containing a Resistive Film{*}}
\author{I.\,M. Braver, Kh.\,L. Garb, and P.\,Sh. Fridberg}
\maketitle
\begin{abstract}
This paper considers the eigenmode problem for a waveguide with a
resistive film. An integral equation for the surface current on the
film is formulated. A method for evaluating the kernel is described,
and an explicit expression for the kernel is derived for rectangular
and circular waveguides.

{*}Originally published in\textit{ Soviet Journal of Communications
Technology and Electronics, vol. 31, no. 2, pp. 33--39, February
1986.} 
\end{abstract}

\renewcommand{\thesection}{\arabic{section}}

\section{DERIVATION OF THE INTEGRAL EQUATION}

\thispagestyle{empty} Consider a uniform waveguide containing an
infinitely thin resistive film with surface impedance $W$. The cross-section
of this waveguide in $z=0$ plane is shown in Fig.~1.

\begin{figure}[H]
\begin{centering}
\includegraphics{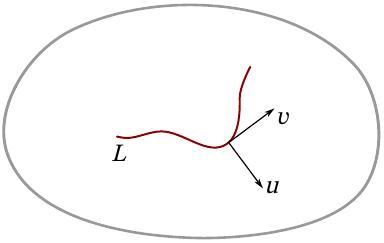}
\par\end{centering}
\caption{Cross-section of an arbitrary-shaped waveguide with a resistive film.}
\end{figure}

Finding the electromagnetic field 
\begin{equation}
\left\{ \begin{array}{c}
\mathfrak{E}\\
\mathfrak{H}
\end{array}\right\} =\left\{ \begin{array}{c}
\mathbf{E}(\mathbf{r})\\
\mathbf{H}(\mathbf{r})
\end{array}\right\} \exp[-i(\omega t-\gamma z)]
\end{equation}
of the eigenmodes reduces to solving the system of wave equations
\begin{equation}
\left(\Delta+\varkappa^{2}\right)F_{p}(\mathbf{r})=0,\qquad F_{p}=\begin{cases}
E_{z}, & p=e,\\
H_{z}, & p=h,
\end{cases}
\end{equation}
where $\mathbf{r}$ is the radius vector in the plane $z=0$, $\Delta$
is the two-dimensional Laplace operator, while $\gamma$ and $\varkappa$
are the propagation constant and transverse wavenumber in the waveguide
with the film. The field components must satisfy (i) the standard
boundary conditions on the perfectly conducting waveguide surface,
(ii) the two-sided boundary conditions on the resistive film \cite{Senior}:
\begin{equation}
\begin{split} & E_{z}(-0,v)=E_{z}(+0,v)=-Wh_{v}(v),\\
 & E_{v}(-0,v)=E_{v}(+0,v)=Wh_{z}(v),
\end{split}
\qquad v\in L,
\end{equation}
where
\begin{equation}
\mathbf{h}(v)\equiv\mathbf{H}(-0,v)-\mathbf{H}(+0,v),
\end{equation}
and (iii) the edge conditions for the resistive film \cite{Braver1988,Braver2024}.
Recall that $h_{u}\equiv0$, while $\left[\mathbf{a}_{u}\times\mathbf{h}\right]$
is the surface-current density on the film; $\mathbf{a}_{u}$ is the
unit vector normal to the film.

To formulate the integral equation, we express $E_{z}$ and $E_{v}$
in terms of the vector $\mathbf{h}$. To this end, we introduce the
Green’s functions $G_{p}$, which satisfy the equation
\begin{equation}
\left(\Delta+\varkappa^{2}\right)G_{p}(\mathbf{r},\mathbf{r}_{1})=-\delta(\mathbf{r}-\mathbf{r}_{1})
\end{equation}
and homogeneous Dirichlet boundary conditions ($p=e$) or Neumann
boundary conditions ($p=h$) on the contour of the waveguide cross-section;
$\delta(\mathbf{r})$ is the Dirac delta function. Using the two-dimensional
Green’s theorem and the boundary conditions on the waveguide surface,
we find
\begin{equation}
\begin{split}F_{p}(\mathbf{r}) & =\underset{L}{\int}dv_{1}\Bigg\{\left[G_{p}(\mathbf{r},\mathbf{r}_{1})\frac{\partial F_{p}(\mathbf{r}_{1})}{\partial u_{1}}-\frac{\partial G_{p}(\mathbf{r},\mathbf{r}_{1})}{\partial u_{1}}F_{p}(\mathbf{r}_{1})\right]_{u_{1}=-0}\\
 & -\left[G_{p}(\mathbf{r},\mathbf{r}_{1})\frac{\partial F_{p}(\mathbf{r}_{1})}{\partial u_{1}}-\frac{\partial G_{p}(\mathbf{r},\mathbf{r}_{1})}{\partial u_{1}}F_{p}(\mathbf{r}_{1})\right]_{u_{1}=+0}\Bigg\}.
\end{split}
\end{equation}

The obtained representation $F_{p}(\mathbf{r})$ is valid only for
observation points outside the film ($\mathbf{r}\notin L$). For these
points, the following relations hold:
\begin{equation}
\begin{split}\left.G_{p}(\mathbf{r},\mathbf{r}_{1})\right|_{u_{1}=-0} & =\left.G_{p}(\mathbf{r},\mathbf{r}_{1})\right|_{u_{1}=+0},\\
\left.\frac{\partial G_{p}(\mathbf{r},\mathbf{r}_{1})}{\partial u_{1}}\right|_{u_{1}=-0} & =\left.\frac{\partial G_{p}(\mathbf{r},\mathbf{r}_{1})}{\partial u_{1}}\right|_{u_{1}=+0}.
\end{split}
\end{equation}
From Maxwell’s equations, the normal derivatives of $E_{z}$ and $H_{z}$
appearing in (6) are found to be 
\begin{equation}
\begin{split}{\displaystyle \frac{\partial E_{z}(\mathbf{r}_{1})}{\partial u_{1}}} & ={\displaystyle -}{\displaystyle \frac{\zeta}{k}}\left[{\displaystyle i\varkappa^{2}H_{v}(\mathbf{r}_{1})+\gamma\frac{\partial H_{z}(\mathbf{r}_{1})}{\partial v_{1}}}\right],\\
\frac{\partial H_{z}(\mathbf{r}_{1})}{\partial u_{1}} & =\frac{1}{k\zeta}\left[i\varkappa^{2}E_{v}(\mathbf{r}_{1})+\gamma\frac{\partial E_{z}(\mathbf{r}_{1})}{\partial v_{1}}\right],
\end{split}
\end{equation}
where $\zeta$ and $k$ are the wave impedance and wavenumber in free
space. By virtue of (3), (7), and (8), expression (6) is transformed
into
\begin{equation}
{\displaystyle E_{z}(\mathbf{r})=-i\zeta\frac{\varkappa^{2}}{k}\underset{L}{\int}dv_{1}\,h_{v}(v_{1})\left.G_{e}(\mathbf{r},\mathbf{r}_{1})\right|_{u_{1}=0}-\zeta\frac{\gamma}{k}\underset{L}{\int}dv_{1}\,\frac{dh_{z}(v_{1})}{dv_{1}}\left.G_{e}(\mathbf{r},\mathbf{r}_{1})\right|_{u_{1}=0},}
\end{equation}
\begin{equation}
H_{z}(\mathbf{r})=-\underset{L}{\int}dv_{1}\,h_{z}(v_{1})\left.\frac{\partial G_{h}(\mathbf{r},\mathbf{r}_{1})}{\partial u_{1}}\right|_{u_{1}=0}.
\end{equation}
The component of interest, $E_{v}(\mathbf{r})$, is given by
\begin{equation}
\begin{split}E_{v}(\mathbf{r}) & =\zeta\frac{\gamma}{k}\frac{\partial}{\partial v}\underset{L}{\int}dv_{1}\,h_{v}(v_{1})\left.G_{e}(\mathbf{r},\mathbf{r}_{1})\right|_{u_{1}=0}-\frac{i\zeta\gamma^{2}}{k\varkappa^{2}}\frac{\partial}{\partial v}\underset{L}{\int}dv_{1}\,\frac{dh_{z}(v_{1})}{dv_{1}}\left.G_{e}(\mathbf{r},\mathbf{r}_{1})\right|_{u_{1}=0}\\
 & +\frac{i\zeta k}{\varkappa^{2}}\frac{\partial}{\partial u}\underset{L}{\int}dv_{1}\,h_{z}(v_{1})\left.\frac{\partial G_{h}(\mathbf{r},\mathbf{r}_{1})}{\partial u_{1}}\right|_{u_{1}=0}.
\end{split}
\end{equation}
Since it is sufficient to know $E_{v}$ on the film to formulate the
integral equation, we do not specify the orientation of the coordinate
frame $(u,v)$ outside the contour $L$. Owning to the continuity,
for ${\displaystyle \mathbf{r}\notin L}$, of the functions ${\displaystyle \left.G_{p}(\mathbf{r},\mathbf{r}_{1})\right|_{u_{1}=0}}$
and ${\displaystyle \left.\frac{\partial G_{h}(\mathbf{r},\mathbf{r}_{1})}{\partial u_{1}}\right|}_{u_{1}=0}$
, in (11) we move differentiation with respect to $u$ and $v$ under
the integral sign. Then, we integrate by parts in (9) and (11) the
terms containing the derivative ${\displaystyle \frac{dh_{z}(v_{1})}{dv_{1}}}$.
We must take into account that for an isolated film, $h_{z}=0$ at
its edges \cite{Braver1988,Braver2024}, while $G_{e}=0$ at the points
of contact for a film touching the waveguide surface. Hence, we obtain
the following expressions:
\begin{equation}
{\displaystyle {\displaystyle E_{z}(\mathbf{r})=-i\zeta\frac{\varkappa^{2}}{k}\underset{L}{\int}dv_{1}\,h_{v}(v_{1})\left.G_{e}(\mathbf{r},\mathbf{r}_{1})\right|_{u_{1}=0}+\zeta\frac{\gamma}{k}\underset{L}{\int}dv_{1}\,h_{z}(v_{1})\left.\frac{\partial G_{e}(\mathbf{r},\mathbf{r}_{1})}{\partial v_{1}}\right|_{u_{1}=0},}}
\end{equation}
\begin{equation}
\begin{split}E_{v}(\mathbf{r}) & =\zeta\frac{\gamma}{k}\underset{L}{\int}dv_{1}\,h_{v}(v_{1})\left.\frac{\partial G_{e}(\mathbf{r},\mathbf{r}_{1})}{\partial v}\right|_{u_{1}=0}+\frac{i\zeta\gamma^{2}}{k\varkappa^{2}}\underset{L}{\int}dv_{1}\,h_{z}(v_{1})\left.\frac{\partial^{2}G_{e}(\mathbf{r},\mathbf{r}_{1})}{\partial v\partial v_{1}}\right|_{u_{1}=0}\\
 & +\frac{i\zeta k}{\varkappa^{2}}\underset{L}{\int}dv_{1}\,h_{z}(v_{1})\left.\frac{\partial^{2}G_{h}(\mathbf{r},\mathbf{r}_{1})}{\partial u\partial u_{1}}\right|_{u_{1}=0}.
\end{split}
\end{equation}
Expressions (12) and (13) were derived under the assumption that $\mathbf{r}\notin L$
($u\neq0$). However, because the components $E_{z}$ and $E_{v}$
are continuous, their values on the film surface can be obtained from
(12) and (13) by taking the limit $u\rightarrow+0$ or $u\rightarrow-0$.
In all terms except $\left.{\displaystyle \frac{\partial^{2}G_{h}(\mathbf{r},\mathbf{r}_{1})}{\partial u\partial u_{1}}}\right|_{u_{1}=0}$,
one can directly set $u=0$. Substituting ${\displaystyle E_{z}(\pm0,v)}$
and ${\displaystyle E_{v}(\pm0,v)}$ into the boundary conditions
(3) yields the vector integral equation for $\mathbf{h}(v)$:
\begin{equation}
\underset{L}{\int}dv_{1}\,\overleftrightarrow{R}(v,v_{1})\cdot\mathbf{h}(v_{1})+W\mathbf{h}(v)=0.
\end{equation}
The components of the tensor ${\displaystyle \overleftrightarrow{R}(v,v_{1})}$
are
\begin{equation}
{\displaystyle \begin{split}R_{vv}(v,v_{1}) & =-i\zeta\frac{\varkappa^{2}}{k}\left.G_{e}(\mathbf{r},\mathbf{r}_{1})\right|_{u=u_{1}=0},\\
R_{vz}(v,v_{1}) & =\zeta\frac{\gamma}{k}\left.\frac{\partial G_{e}(\mathbf{r},\mathbf{r}_{1})}{\partial v_{1}}\right|_{u=u_{1}=0},\\
R_{zv}(v,v_{1}) & =-\zeta\frac{\gamma}{k}\left.\frac{\partial G_{e}(\mathbf{r},\mathbf{r}_{1})}{\partial v}\right|_{u=u_{1}=0},\\
R_{zz}(v,v_{1}) & =-i\zeta\frac{k}{\varkappa^{2}}\left\{ \left.\frac{\gamma^{2}}{k^{2}}\frac{\partial^{2}G_{e}(\mathbf{r},\mathbf{r}_{1})}{\partial v\partial v_{1}}\right|_{u=u_{1}=0}+D(v,v_{1})\right\} ,
\end{split}
}
\end{equation}
where
\begin{equation}
D(v,v_{1})=\lim_{u\to\pm0}\frac{\partial}{\partial u}\left[\left.\frac{\partial G_{h}(\mathbf{r},\mathbf{r}_{1})}{\partial u_{1}}\right|_{u_{1}=0}\right].
\end{equation}
The solution of equation (14) should be sought for in the class of
functions satisfying the edge conditions \cite{Braver1988,Braver2024}.

To complete the formulation of the integral equation (14), we must
specify how to calculate the limit (16). For problems with translational
symmetry along the coordinate $u$ (that is, when ${\displaystyle G}_{h}$
depends on $\left|{\displaystyle u-u_{1}}\right|$), the calculation
of the limit is described in \cite{Honl}. The absence of this symmetry
in the present problem prevents us from applying the results of \cite{Honl}.
Another method for calculating (16) may be applied in the case the
function ${\displaystyle G}_{h}$ represented as a series of eigenfunctions.
Term-by-term differentiation of this series is allowed only if the
resulting series is subsequently summed in the Abel--Poisson sense.
This procedure is readily implemented for problems solved in Cartesian
coordinates. In other cases (for example, when the eigenfunctions
depend on polar coordinates), implementing this procedure poses considerable
difficulties. In this paper, methods are proposed for calculating
the limit (16) for rectangular and circular waveguides without summing
the series in the Abel--Poisson sense.

\section{RECTANGULAR WAVEGUIDE WITH A RESISTIVE FILM (Fig. 2)}

\begin{figure}[H]
\begin{centering}
\includegraphics{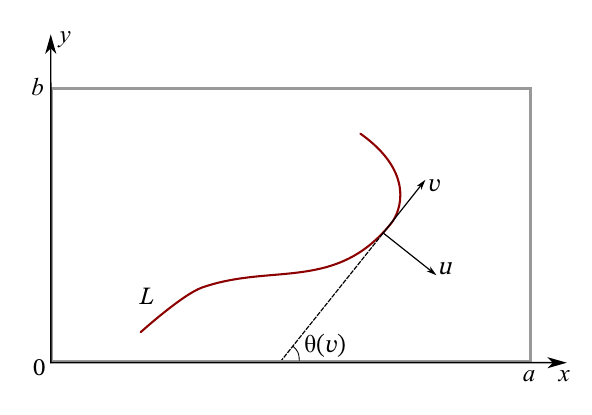}
\par\end{centering}
\caption{Cross-section of a rectangular waveguide with a resistive film.}
\end{figure}

We introduce four Green functions satisfying various boundary conditions
on the waveguide surface:
\begin{equation}
\Gamma_{pq}(x,y;x_{1},y_{1})=\sum^{\infty}_{m,\,n=0}\frac{\psi_{pm}(a\mid x,x_{1})\,\psi_{qn}(b\mid y,y_{1})}{\alpha^{2}_{m}+\beta^{2}_{n}-\varkappa^{2}},
\end{equation}
\begin{equation}
\psi_{pm}(a\mid x,x_{1})=\frac{\varepsilon_{m}}{a}\begin{cases}
\sin\left(\dfrac{m\pi}{a}x\right)\,\sin\left(\dfrac{m\pi}{a}x_{1}\right), & p=e,\\[6pt]
\cos\left(\dfrac{m\pi}{a}x\right)\,\cos\left(\dfrac{m\pi}{a}x_{1}\right), & p=h,
\end{cases}
\end{equation}
${\displaystyle q=e,h}$; ${\displaystyle \alpha_{m}=\frac{m\pi}{a}}$,
${\displaystyle \beta_{n}=\frac{n\pi}{b}}$; ${\displaystyle \varepsilon_{m}=2-\delta_{m0}}$,
$\delta_{mn}$ is the Kronecker symbol. Clearly, the functions $G_{P}$
introduced in Section 1 coincide with $\Gamma_{pp}$ (i.e., ${\displaystyle G_{p}\equiv\Gamma_{pp}}$).
Direct differentiation of (17) readily yields the following identity:
\begin{equation}
\frac{\partial^{2}G_{h}}{\partial u\partial u_{1}}+\frac{\partial^{2}G_{e}}{\partial v\partial v_{1}}=-\cos\theta(v)\cos\theta(v_{1})\,\Delta\Gamma_{he}-\sin\theta(v)\sin\theta(v_{1})\,\Delta\Gamma_{eh}.
\end{equation}
Since the functions $\Gamma_{pq}$ satisfy the differential equation
(5), the identity
\begin{equation}
\lim_{u\to\pm0}\left[\delta(\mathbf{r}-\mathbf{r}_{1})\right]_{u_{1}=0}=0
\end{equation}
gives the following expression: 
\begin{equation}
\begin{aligned}D(v,v_{1})=\Bigg[ & \varkappa^{2}\cos\theta(v)\cos\theta(v_{1})\,\Gamma_{he}\\
 & +\varkappa^{2}\sin\theta(v)\sin\theta(v_{1})\,\Gamma_{eh}-\frac{\partial^{2}G_{e}}{\partial v\partial v_{1}}\Bigg]_{u=u_{1}=0}.
\end{aligned}
\end{equation}
Special cases of expression (21) for ${\displaystyle \theta(v)\equiv\frac{\pi}{2}}$
(a film in the $E$-plane) and ${\displaystyle \theta(v)=\arctan\frac{b}{a}}$
(a film in the diagonal plane) were obtained earlier \cite{Braver1985,Braver1984}
by summing the series in the Abel--Poisson sense.

\section{CIRCULAR WAVEGUIDE WITH A RESISTIVE FILM (Fig. 3)}

\begin{figure}[H]
\begin{centering}
\includegraphics{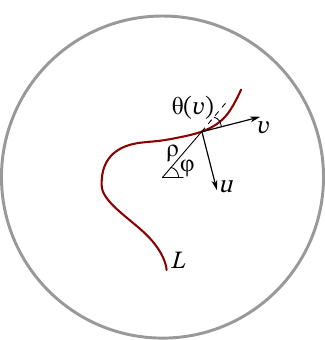} 
\par\end{centering}
\caption{Cross-section of a circular waveguide with a resistive film.}
\end{figure}

The Green’s functions of a circular waveguide of unit radius have
the form 
\begin{equation}
G_{p}(\rho,\varphi;\rho_{1},\varphi_{1})=\sum^{\infty}_{\substack{m=0\\
n=1
}
}\frac{\varepsilon_{m}}{\pi}\frac{\psi_{pmn}(\rho,\rho_{1})\cos m(\varphi-\varphi_{1})}{\varkappa^{2}_{pmn}-\varkappa^{2}}-\frac{1}{\pi\varkappa^{2}}\psi_{p00},
\end{equation}
\begin{equation}
\psi_{pmn}(\rho,\rho_{1})=C_{pmn}J_{m}(\varkappa_{pmn}\rho)J_{m}(\varkappa_{pmn}\rho_{1}),\qquad\psi_{e00}=0,\qquad\psi_{h00}=1,
\end{equation}
\begin{equation}
C_{pmn}=\begin{cases}
J^{-2}_{m+1}(\varkappa_{emn}), & p=e,\\[6pt]
\dfrac{\varkappa^{2}_{hmn}}{\varkappa^{2}_{hmn}-m^{2}}J^{-2}_{m}(\varkappa_{hmn}), & p=h,
\end{cases}
\end{equation}
where ${\displaystyle {\displaystyle \varkappa}_{pmn}}$ are the positive
roots of the equations $J_{m}(\varkappa_{emn})=0$ and $J^{\prime}_{m}(\varkappa_{hmn})=0$;
$J_{m}(x)$ and $J^{\prime}_{m}(x)$ are the Bessel function and its
derivative. Summing (22) over $n$ (see the Appendix), we obtain another
representation 
\begin{equation}
G_{p}(\rho,\varphi;\rho_{1},\varphi_{1})=\frac{1}{4}\sum^{\infty}_{m=0}\varepsilon_{m}g_{pm}(\rho,\rho_{1})\cos m(\varphi-\varphi_{1}),
\end{equation}
where 
\begin{equation}
g_{pm}(\rho,\rho_{1})=\begin{cases}
J_{m}(\varkappa\rho)\left[J_{m}(\varkappa\rho_{1})X_{pm}(\varkappa)-Y_{m}(\varkappa\rho_{1})\right], & 0\leq\rho\leq\rho_{1}\leq1,\\[6pt]
J_{m}(\varkappa\rho_{1})\left[J_{m}(\varkappa\rho)X_{pm}(\varkappa)-Y_{m}(\varkappa\rho)\right], & 0\leq\rho_{1}\leq\rho\leq1,
\end{cases}
\end{equation}
\begin{equation}
X_{pm}(\varkappa)=\begin{cases}
\frac{Y_{m}(\varkappa)}{J_{m}(\varkappa)}, & p=e,\\[4pt]
\frac{Y^{\prime}_{m}(\varkappa)}{J^{\prime}_{m}(\varkappa)}, & p=h,
\end{cases}
\end{equation}
with ${\displaystyle Y_{m}(x)}$ and ${\displaystyle Y^{\prime}_{m}(x)}$
being the Neumann function and its derivative. For an arbitrary contour
$L$ (except when the film lies in the diametral plane), the derivative
${\displaystyle \frac{\partial}{\partial u}}$ can be expressed in
terms of $\frac{{\displaystyle \partial}}{{\displaystyle \partial\rho}}$
and $\frac{{\displaystyle \partial}}{{\displaystyle \partial v}}$:
\begin{equation}
\begin{aligned}\frac{\partial^{2}G_{h}}{\partial u\partial u_{1}}= & \Bigg[\cos\theta(v)\cos\theta(v_{1})\frac{\partial^{2}G_{h}}{\partial v\partial v_{1}}-\cos\theta(v)\frac{\partial^{2}G_{h}}{\partial v\partial\rho_{1}}\\
 & \quad-\cos\theta(v_{1})\frac{\partial^{2}G_{h}}{\partial\rho\partial v_{1}}+\frac{\partial^{2}G_{h}}{\partial\rho\partial\rho_{1}}\Bigg]\csc\theta(v)\csc\theta(v_{1}).
\end{aligned}
\end{equation}
It follows that evaluating (16) reduces to calculating the following
limit: 
\begin{equation}
\lim_{u\to\pm0}\frac{\partial}{\partial\rho}\left(\left.\frac{\partial G_{h}}{\partial\rho_{1}}\right|_{u_{1}=0}\right).
\end{equation}
We now derive its explicit expression. Differentiating ${\displaystyle G_{h}}$
with respect to $\rho$ and $\rho_{1}$ requires differentiating the
functions $g_{hm}$. At $\rho=\rho_{1}$, these functions have a kink,
so ${\displaystyle \frac{\partial g_{hm}}{\partial\rho_{1}}}$ has
a jump discontinuity, while ${\displaystyle \frac{\partial^{2}g_{hm}}{\partial\rho\partial\rho_{1}}}$
contains the Dirac delta function ${\displaystyle \delta(\rho-\rho_{1})}$.
Since in (29) ${\displaystyle {\displaystyle u\neq u_{1}}}$, and
therefore $\rho\neq\rho_{1}$, the following expression is obtained:
\begin{equation}
{\displaystyle \lim_{u\to\pm0}\frac{\partial}{\partial\rho}\left(\left.\frac{\partial g_{hm}}{\partial\rho_{1}}\right|_{u_{1}=0}\right)=\varkappa^{2}\begin{cases}
J^{\prime}_{m}(\varkappa\rho)\left[J^{\prime}_{m}(\varkappa\rho_{1})X_{hm}(\varkappa)-Y^{\prime}_{m}(\varkappa\rho_{1})\right], & 0\leq\rho\leq\rho_{1}\leq1,\\[6pt]
J^{\prime}_{m}(\varkappa\rho_{1})\left[J^{\prime}_{m}(\varkappa\rho)X_{hm}(\varkappa)-Y^{\prime}_{m}(\varkappa\rho)\right], & 0\leq\rho_{1}\leq\rho\leq1.
\end{cases}}
\end{equation}
Formulas (25), (28), and (30) allow us to write down ${\displaystyle D(v,v_{1})}$
explicitly. For example, when ${\displaystyle \theta(v)\equiv\frac{\pi}{2}}$
and ${\displaystyle v=b\varphi}$ (the film is an arc of a concentric
circle of radius ${\displaystyle b<1}$), we obtain the following
expression: 
\begin{equation}
D(v,v_{1})=\frac{\varkappa^{2}}{4}\sum^{\infty}_{m=0}\varepsilon_{m}J^{\prime}_{m}(\varkappa b)\left[J^{\prime}_{m}(\varkappa b)X_{hm}(\varkappa)-Y^{\prime}_{m}(\varkappa b)\right]\cos m(\varphi-\varphi_{1}).
\end{equation}
We now return to the previously excluded case of a film in the diametral
plane. Without loss of generality, we may assume that the film lies
in the plane ${\displaystyle \varphi=0,\,\pi}$ (the diametral plane).
The resulting relation is: 
\begin{equation}
\frac{\partial}{\partial u}=\sigma(v)\frac{\partial}{\rho\partial\varphi},\qquad\sigma(v)=\begin{cases}
-1, & \varphi=0,\\
+1, & \varphi=\pi.
\end{cases}
\end{equation}
In this problem, ${\displaystyle D(v,v_{1})}$ is determined through
${\displaystyle \frac{\partial^{2}G_{h}}{\partial\varphi\partial\varphi_{1}}}$.
Taking into account that 
\begin{equation}
\frac{\partial^{2}G_{h}}{\partial\varphi\partial\varphi_{1}}=-\frac{\partial^{2}G_{h}}{\partial\varphi^{2}}
\end{equation}
and using the differential equation for the function $G_{h}$,
\begin{equation}
\frac{1}{\rho}\frac{\partial}{\partial\rho}\left(\rho\frac{\partial G_{h}}{\partial\rho}\right)+\frac{1}{\rho^{2}}\frac{\partial^{2}G_{h}}{\partial\varphi^{2}}+\varkappa^{2}G_{h}=-\frac{1}{\rho}\delta(\rho-\rho_{1})\delta(\varphi-\varphi_{1}),
\end{equation}
the resulting expression is
\begin{equation}
D(v,v_{1})=\sigma(v)\sigma(v_{1})\frac{\rho}{\rho_{1}}\left[\frac{1}{\rho}\frac{\partial}{\partial\rho}\left(\rho\frac{\partial G_{h}}{\partial\rho}\right)+\varkappa^{2}G_{h}\right]_{u=u_{1}=0}.
\end{equation}
Thus, the expressions obtained for ${\displaystyle D(v,v_{1})}$ cover
all possible configurations of the film in a circular waveguide.

\appendix
\setcounter{equation}{0}
\renewcommand{\theequation}{A.\arabic{equation}} 

\section{APPENDIX}

The following identities are used in the derivation of (25): 
\begin{equation}
\sum^{\infty}_{n=1}\frac{J_{m}(\varkappa_{emn}\rho)J_{m}(\varkappa_{emn}x)}{\left(\varkappa^{\,2}_{emn}-\varkappa^{2}\right)J^{2}_{m+1}(\varkappa_{emn})}=\frac{\pi}{4}g_{em}(\rho,x),\qquad m=0,1,\ldots
\end{equation}
\begin{equation}
-\frac{\delta_{m0}}{\varkappa^{2}}+\sum^{\infty}_{n=1}\frac{\varkappa^{2}_{hmn}}{\varkappa^{2}_{hmn}-m^{2}}\cdot\frac{J_{m}(\varkappa_{hmn}\rho)J_{m}(\varkappa_{hmn}x)}{\left(\varkappa^{2}_{hmn}-\varkappa^{2}\right)J^{2}_{m}(\varkappa_{hmn})}=\frac{\pi}{4}g_{hm}(\rho,x),\qquad m=0,1,\ldots
\end{equation}
To prove (A.1) and (A.2), we use the Fourier--Bessel expansion \cite{Watson}
of the function ${\displaystyle {\displaystyle g_{em}(\rho,x)}}$
and the Dini expansion of the function $g_{hm}(\rho,x)$: 
\begin{equation}
g_{pm}(\rho,x)=a_{p00}\delta_{m0}+\sum^{\infty}_{n=1}a_{pmn}J_{m}(\varkappa_{pmn}x),
\end{equation}
\begin{equation}
a_{pmn}=2C_{pmn}\intop^{1}_{0}J_{m}(\varkappa_{pmn}\xi)\,g_{pm}(\rho,\xi)\,\xi\,d\xi,\qquad a_{e00}=0,\quad a_{h00}=2\intop^{1}_{0}g_{h0}(\rho,\xi)\,\xi\,d\xi.
\end{equation}
Integrating with respect to $\xi$ and using the Wronskian identity
\begin{equation}
J_{m}(x)Y^{\prime}_{m}(x)-J^{\prime}_{m}(x)Y_{m}(x)=\frac{2}{\pi x},
\end{equation}
we find 
\begin{equation}
\begin{split} & \intop^{1}_{0}J_{m}(\varkappa_{pmn}\xi)\,g_{pm}(\rho,\xi)\,\xi\,d\xi=\frac{2}{\pi}\frac{J_{m}(\varkappa_{pmn}\rho)}{\varkappa^{2}_{pmn}-\varkappa^{2}},\\
 & \intop^{1}_{0}g_{h0}(\rho,\xi)\,\xi\,d\xi=-\frac{2}{\pi\varkappa^{2}}.
\end{split}
\end{equation}
Substituting the integrals (A.6) into (A.3), we arrive at (A.1) and
(A.2).

We include the derivation of (A.1) because the analogous formulas
in \cite[Chapter 15.42]{Watson} and \cite[Chapter 7.15, Equation (58)]{Bateman}
do not coincide. Equation (A.1) agrees with the result obtained in
\cite{Bateman}. The expansion (A.2) does not appear in the works
known to us.

\pagebreak{}

\end{document}